\documentclass{PoS}

\usepackage{xspace}
\usepackage{amsmath}
\usepackage{multicol}
\usepackage{overpic}
\usepackage{xcolor}

\newcommand{\sn}[2]{\ensuremath{#1\times 10^{#2}}\xspace}

\newcommand{\mpsr}{\ensuremath{M_{\text{psr}}}\xspace}
\newcommand{\mco}{\ensuremath{M_{\text{co}}}\xspace}
\newcommand{\mbe}{\ensuremath{M_{\text{Be}}}\xspace}
\newcommand{\rbe}{\ensuremath{R_{\text{Be}}}\xspace}

\newcommand{\tbe}{\ensuremath{T_{\text{Be}}}\xspace}
\newcommand{\mdot}{\ensuremath{\dot{M}_{\text{w}}}\xspace}
\newcommand{\vw}{\ensuremath{v_{\text{w}}}\xspace}
\newcommand{\bestar}{MWC 148\xspace}
\newcommand{\edot}{\ensuremath{L_{\text{sd}}}\xspace}
\newcommand{\thetaic}{\ensuremath{\theta_{\text{ICS}}}\xspace}
\newcommand{\porb}{\ensuremath{P_{\text{orb}}}\xspace}
\newcommand{\rsh}{\ensuremath{R_{\text{sh}}}\xspace}

\newcommand{\norm}{\ensuremath{N_e}\xspace}
\newcommand{\slope}{\ensuremath{\Gamma}\xspace}
\newcommand{\normi}[1]{\ensuremath{N_{e,#1}}\xspace}
\newcommand{\slopei}[1]{\ensuremath{\Gamma_{#1}}\xspace}

\newcommand{\msun}{\ensuremath{M_{\odot}}\xspace}
\newcommand{\msunyr}{\ensuremath{M_{\odot}\text{/yr}}\xspace}
\newcommand{\rsun}{\ensuremath{R_{\odot}}\xspace}
\newcommand{\dg}{\ensuremath{^{\circ}}\xspace}

\newcommand{\src}{HESS~J0632$+$057\xspace}
\newcommand{\nustar}{{\it NuSTAR}\xspace}

\newcommand{\fermi}{{\it Fermi-}LAT\xspace}

\newcommand{\asec}{\ensuremath{\prime\prime}\xspace}

\newcommand{\flux}{erg\,cm$^{-2}$\,s$^{-1}$\xspace}
\newcommand{\lum}{erg\,s$^{-1}$\xspace}

\newcommand{\apj}{ApJ}
\newcommand{\aap}{A\&A}
\newcommand{\mnras}{Monthly Notices of the RAS}

\title{Combined VERITAS and NuSTAR observations of the gamma-ray binary \src}

\ShortTitle{NuSTAR and VERITAS observations of \src}

\author{
\speaker{Raul R. Prado}$^{a}$,
Charles Hailey$^{b}$, Shifra Mandel$^{b}$, Kaya Mori$^{b}$,
for the NuSTAR\footnote{https://www.nustar.caltech.edu/} and VERITAS\footnote{https://veritas.sao.arizona.edu/} Collaborations\\
\llap{$^a$} Deutsches Elektronen-Synchrotron (DESY), Platanenallee 6, 15738 Zeuthen, Germany\\
\llap{$^b$} Columbia Astrophysics Laboratory, Columbia University, New York, NY 10027, USA\\
E-mail: \email{raul.prado@desy.de}
}

\abstract{\src is a gamma-ray binary composed of a compact object and a Be star, with an orbital period of about 315 days. The actual nature of its non-thermal emission, spanning from radio to very-high-energy (VHE, >100 GeV) gamma-rays, is currently unknown. In this contribution we will present the results of a set of simultaneous observations performed by the NuSTAR X-ray telescope and the VERITAS observatory. The combination of hard X-rays (3-30 keV) and VHE gamma-rays (0.1-5 TeV) provide valuable information for the understanding of the radiative processes occurring in the system. The spectral energy distributions (SED) derived from the observations are used to probe the pulsar scenario, in which the system is powered by a rapidly rotating neutron star. The non-thermal emission is produced by the particles accelerated at the shock formed by the collision of the pulsar and stellar winds. As a results of the model fitting, we constrain the relation between the pulsar spin-down luminosity and the magnetization of the pulsar wind.}

\FullConference{36th International Cosmic Ray Conference -ICRC2019-\\
		July 24th - August 1st, 2019\\
		Madison, WI, U.S.A.}

\begin{document}

\section{Introduction}
\label{sec:intro}

\vspace{-2mm}

Gamma-ray binaries are defined as binary systems composed of a compact object and a massive star, in which the spectral energy distribution peaks at the gamma-ray band ($>1$ MeV)~\cite{Dubus-2013}. Only a few sources have been unambiguously identified as gamma-ray binaries: PSR~B1259-63, LS~5039, LS~I$+$61~303, \src, 1FGL~J1018.6$-$5856, LMC P3, PSR~J2032$+$4127~\cite{Paredes-2019} and 3FGL J1405.4-6119~\cite{Corbet-2019}. In all these systems, the massive star is of the O or B type. The orbital period may vary substantially, from 3.9~days (LS~5039) to ${\sim}50$~years (PSR~J2032$+$4127). The nature of the compact object is only known in PSR~B1259$-$63 and PSR~J2032$+$4127, which contain radio pulsars.

\src was first detected by H.E.S.S. during observations of the Monoceros region \cite{Aharonian-2007}. Later observations by VERITAS did not lead to a detection \cite{Acciari-2009}, indicating a substantial flux variability, which is characteristic of gamma-ray binaries. The system has been extensively observed since then in soft X-rays \cite{Falcone-2010} and the TeV gamma-ray band \cite{Aliu-2014, Aleksic-2012}. The light curve resultant from both X-ray and gamma-ray observations presents two clear periodic outbursts around $\phi\approx0.3-0.4$ and $\phi\approx0.6-0.8$ when folded to an orbital period of around $315-320$ days~\cite{Aliu-2014}. In the GeV gamma-ray band, the system is very faint where it was only recently detected in \fermi data~\cite{Li-2017}.  

The orbital period of \src was initially derived to be $\porb=321$ days by \cite{Bongiorno-2011} using X-ray light-curve data, and was later refined to be $\porb=315^{+6}_{-4}$ days~\cite{Aliu-2014}. There are two distinct orbital solutions available in Ref.~\cite{Casares-2012} and \cite{Moritani-2018}. Despite following similar methodologies, both studies resulted in completely different sets of orbital parameters. The compact object in \src is still unknown and the two orbital solutions point in opposite directions. While the solutions from Ref.~\cite{Moritani-2018}  suggests that the mass of the compact object is consistent with a pulsar ($\mco<2.5$ \msun), the solution by~\cite{Casares-2012} favors the black-hole scenario with $\mco > 2.1~ \msun$~\cite{Zamanov-2017}, being only marginally compatible with the pulsar scenario. 

In this contribution, we present contemporaneous observations in X-ray by \nustar\ and TeV gamma-ray by VERITAS during November and December 2017. The observations correspond to orbital phases $\approx0.22$ and $0.30$, respectively. The SEDs are used to probe a model based on the pulsar scenario. The model assumes that the non-thermal emission is produced by electrons from the pulsar wind that are accelerated at the termination shock formed by the collision of the stellar and pulsar wind. Synchrotron radiation and inverse Compton scattering of stellar photons are assumed to be the mechanisms responsible for producing the X-ray and TeV gamma-ray photons observed, respectively. The result of the model fitting provides a relation between the pulsar spin-down luminosity (\edot) and the pulsar-wind magnetization ($\sigma$) that is consistent with theoretical expectations, showing that our data can be satisfactorily described within the pulsar hypothesis.

\vspace{-3mm}

\section{Observations and Data Analysis}
\label{sec:observations}

\vspace{-2mm}

\newcommand{\icrc}{\textcolor{black!50}{ICRC 2019}}

\begin{figure}[!h]
\centering

\begin{overpic}[clip, width=0.40\textwidth]{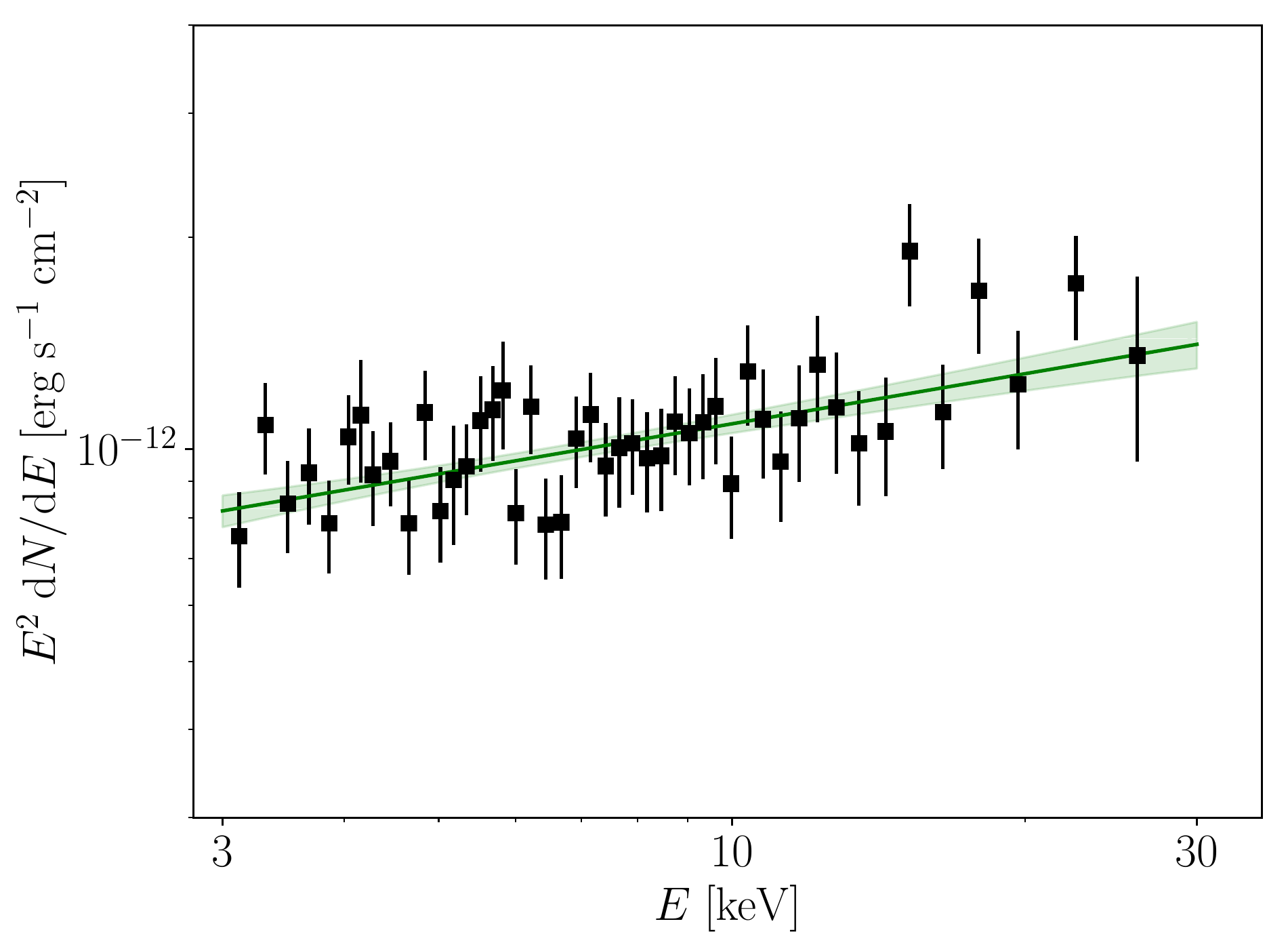}
\put(25,15){\icrc}
\end{overpic}
\hspace{5mm}
\begin{overpic}[clip, width=0.40\textwidth]{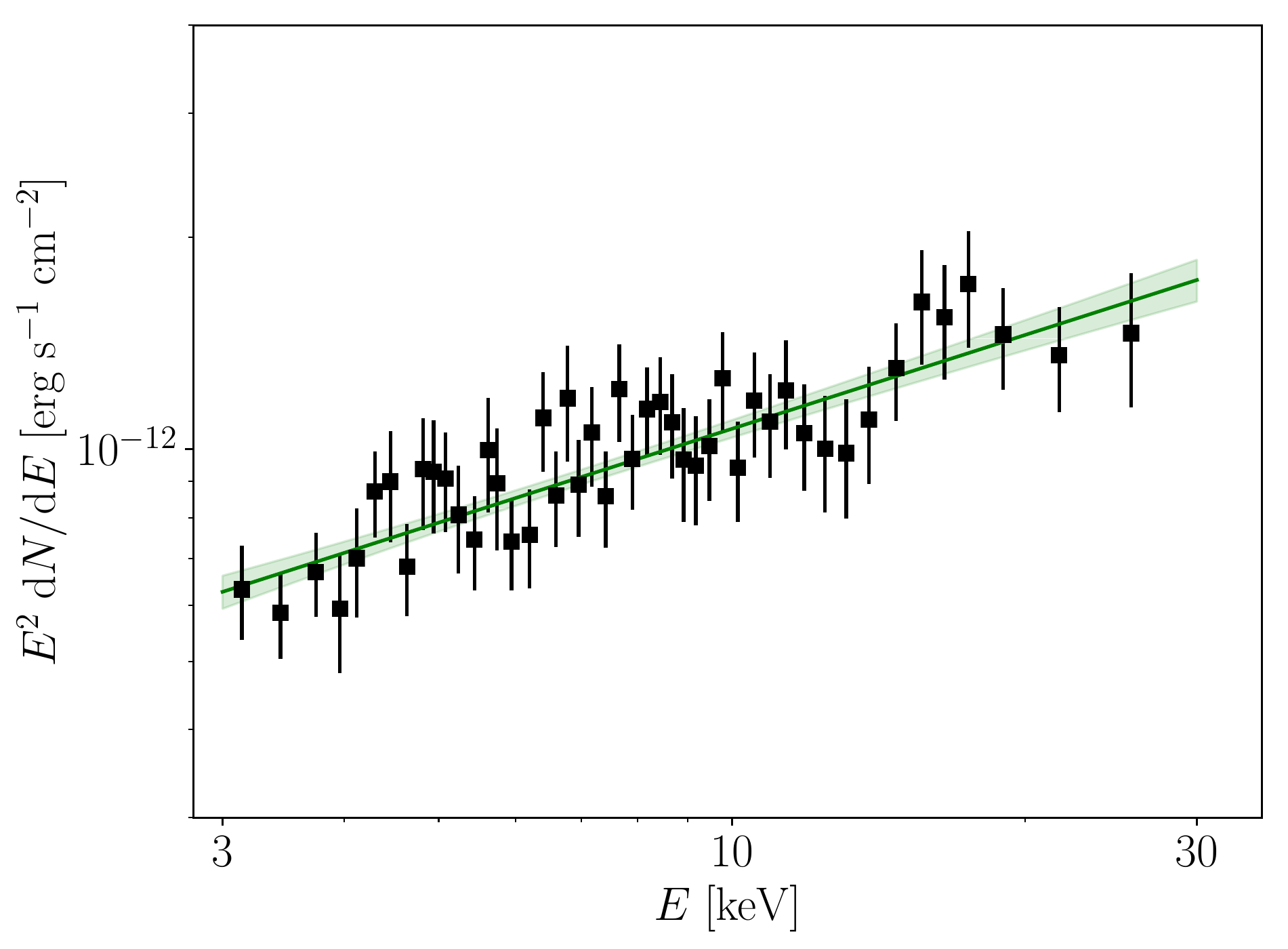}
\put(25,15){\icrc}
\end{overpic}

\begin{overpic}[clip, width=0.40\textwidth]{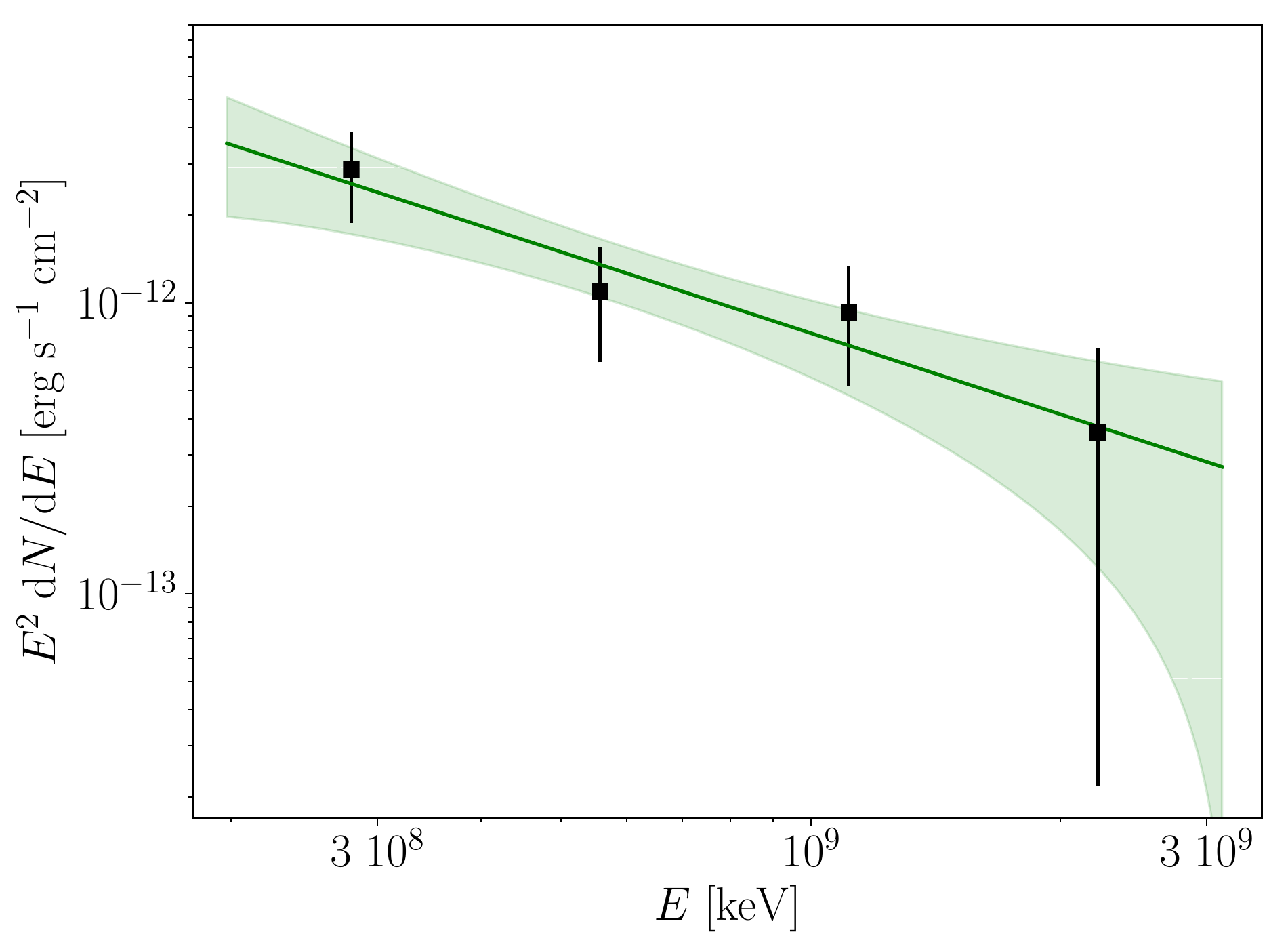}
\put(25,15){\icrc}
\end{overpic}
\hspace{5mm}
\begin{overpic}[clip, width=0.40\textwidth]{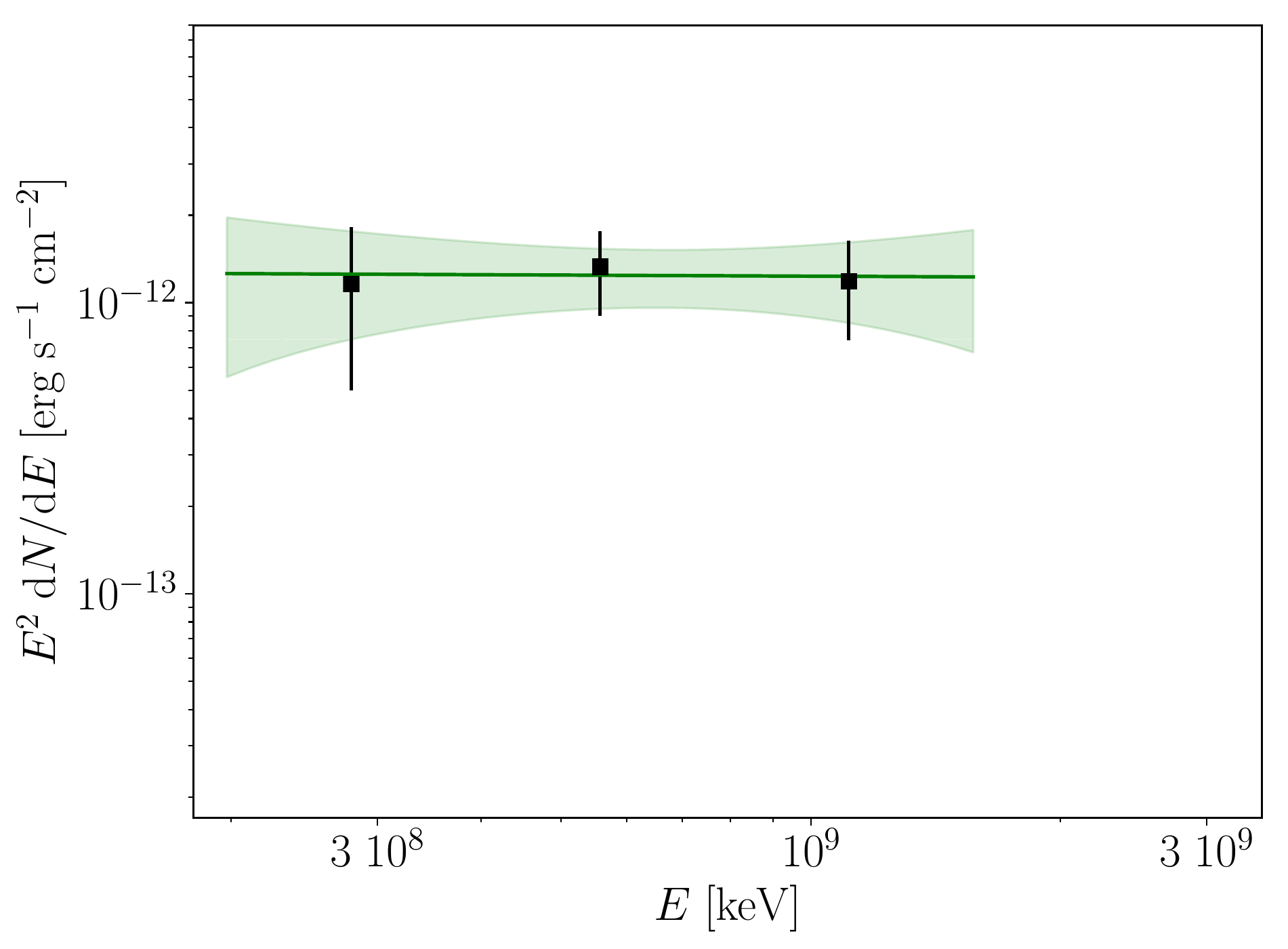}
\put(25,15){\icrc}
\end{overpic}

\caption{SED from \nustar (upper) and VERITAS (lower) observations from November (left) and December (right) 2017. The dashed lines show the result of the single power law fit and the green band its 1$\sigma$ confidence interval.}
\label{fig:data}
\end{figure}

\subsection{NuSTAR}
\label{sec:nustar}

\vspace{-2mm}

\nustar consists of a pair of co-aligned high-energy X-ray focusing telescopes with focal plane modules FPMA and FPMB, providing an imaging resolution of 18\asec\ FWHM over the energy range from 3 to 79 keV, and a characteristic 400 eV FWHM spectral resolution at 10 keV~\cite{Harrison-2013}. \nustar's absolute and relative timing accuracy after correcting for on-board clock drift are 3 msec and 10 $\mu$sec, respectively \cite{Madsen-2015}. The broadband capabilities of \nustar allow us to measure spectral properties such as photon indices with high precision, with little to no dependence on ISM absorption ($N_{\rm H}$).

\src was observed by \nustar on November 22, 2017 and December 14, 2017, with 49.7~ks and 49.6~ks exposures, respectively. The data processing and analysis were completed using the HEASOFT (V6.22) software package, including NUSTARDAS 06Jul17$\_$v1.8.0. \nustar spectra were grouped to a minimum significance of 5$\sigma$ in each energy bin. We fit \nustar module A and B spectra jointly in the 3--30 keV energy band, above which the background dominates. Given the previously measured column density values ($N_{\rm H}\sim (2.1-4.7)\times10^{21}$ cm$^{-2}$) \cite{Moritani-2018}, it was found that the ISM absorption is negligible above 3 keV. Therefore, \nustar spectra allow us to determine the intrinsic continuum spectral index independently, without degeneracy with $N_{\rm H}$. 

In Fig.~\ref{fig:data} (upper plots), we show the SED together with the results of a single power-law fit. For the November observation, the 0.3--30 keV flux, corresponding luminosity and photon index obtained were $(2.42\pm0.13)\times10^{-12}$~\flux, $(5.67\pm0.30)\times10^{32}$~\lum and $\Gamma = 1.77\pm0.05$, respectively. The December spectrum is significantly harder, with photon index $\Gamma = 1.56\pm0.05$, 0.3--30 keV flux $(2.45\pm0.13)\times10^{-12}$~\flux, and luminosity of $(5.75\pm0.30)\times10^{32}$~\lum. Luminosity values assume a distance of 1.4~kpc.

\subsection{VERITAS}
\label{sec:veritas}

\vspace{-2mm}

VERITAS consists of an array of four 12m-diameter telescopes located at the Fred Lawrence Whipple Observatory (FLWO) in southern Arizona (31 40N, 110 57W,  1.3km a.s.l.)~\cite{Weekes-2002}. It is designed to observe gamma-ray sources in the energy range from 100 GeV to $>$30 TeV. Gamma-ray photons are detected through the Cherenkov light induced by the cascade of secondary particles produced after its interaction with the atmosphere. The VERITAS sensitivity enables the detection of a source with 1\% of the Crab flux within approximately 25 hours with an angular resolution of $<0.1\dg$ at 1 TeV~\cite{Park-2015}.

VERITAS observations of \src were conducted over 7.4 hours between November 16 and 26, 2017 and for 6.0 hours between December 14 and 16, 2017. Observations were performed in ``wobble'' mode with 0.5\dg offset from the center of the telescope's field-of-view. The VERITAS data were analyzed following the standard procedure described in Ref.~\cite{Maier-2017}. The images were first calibrated, cleaned and parameterized using the Hillas criteria~\cite{Hillas-1985, Krawczynski-2006, Daniel-2008}. The arrival direction and core location were determined by using a stereoscopic technique that combines the orientation of the images from different telescopes.

The SED derived from the VERITAS data is shown in Fig.~\ref{fig:data} (lower panels), together with the results of a fit of a single power law. The flux in the range $0.2-3$ TeV, the corresponding luminosity, and the power law indices were found to be $(3.47\pm0.81)\times 10^{-12}$~\flux, $(8.14\pm1.90)\times10^{32}$~\lum and $2.93\pm0.49$ for the November observation, and $(3.36\pm0.80)\times10^{-12}$~\flux, $(7.88\pm1.88)\times10^{32}$~\lum and $2.02\pm0.43$ for the December observation. 

\vspace{-3mm}

\section{System Parameters and Orbital Solutions}
\label{sec:system}

\vspace{-2mm}

\begin{figure}[t!]
\centering
\includegraphics[width=0.40\textwidth]{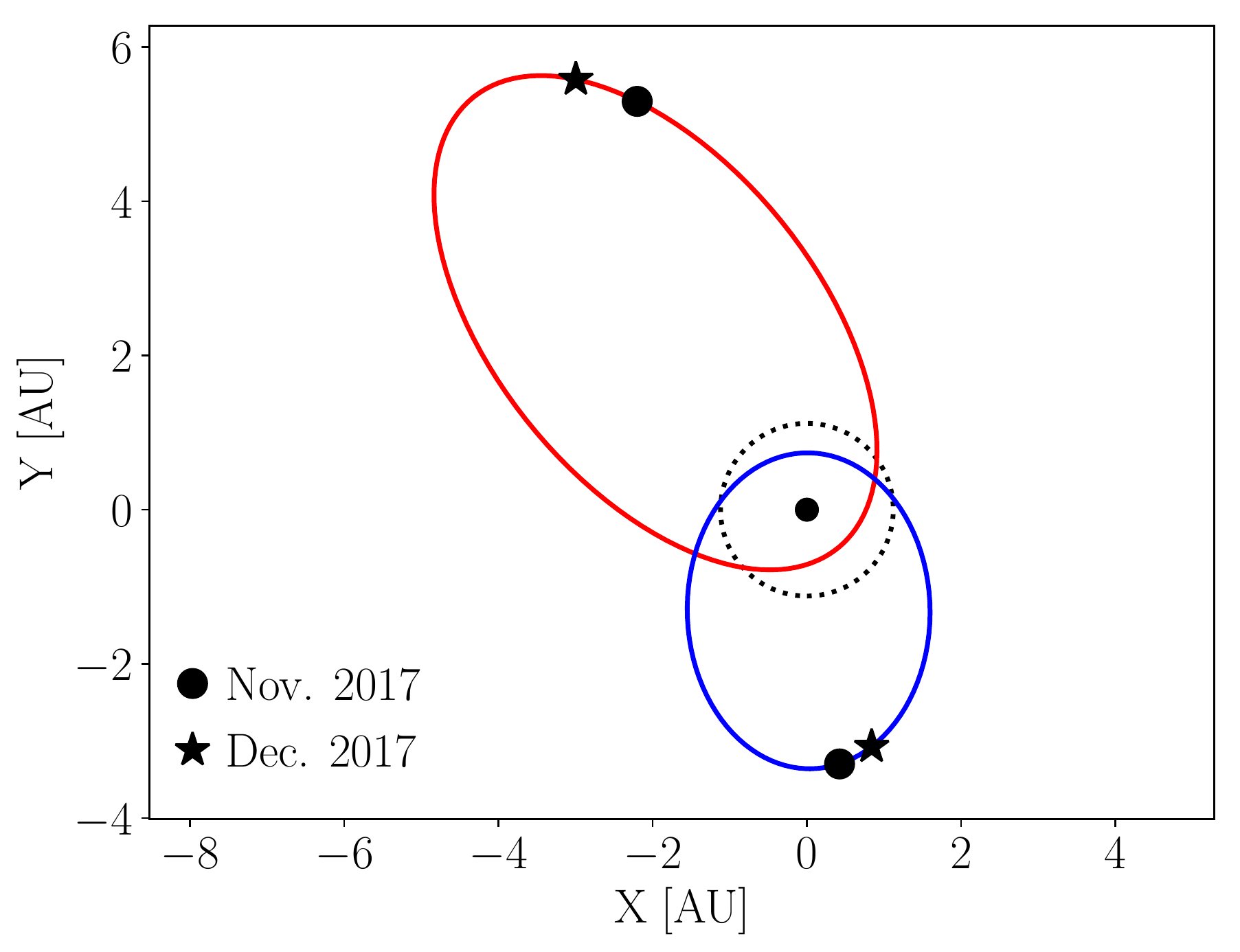}
\hspace{8mm}
\begin{overpic}[clip, width=0.43\textwidth]{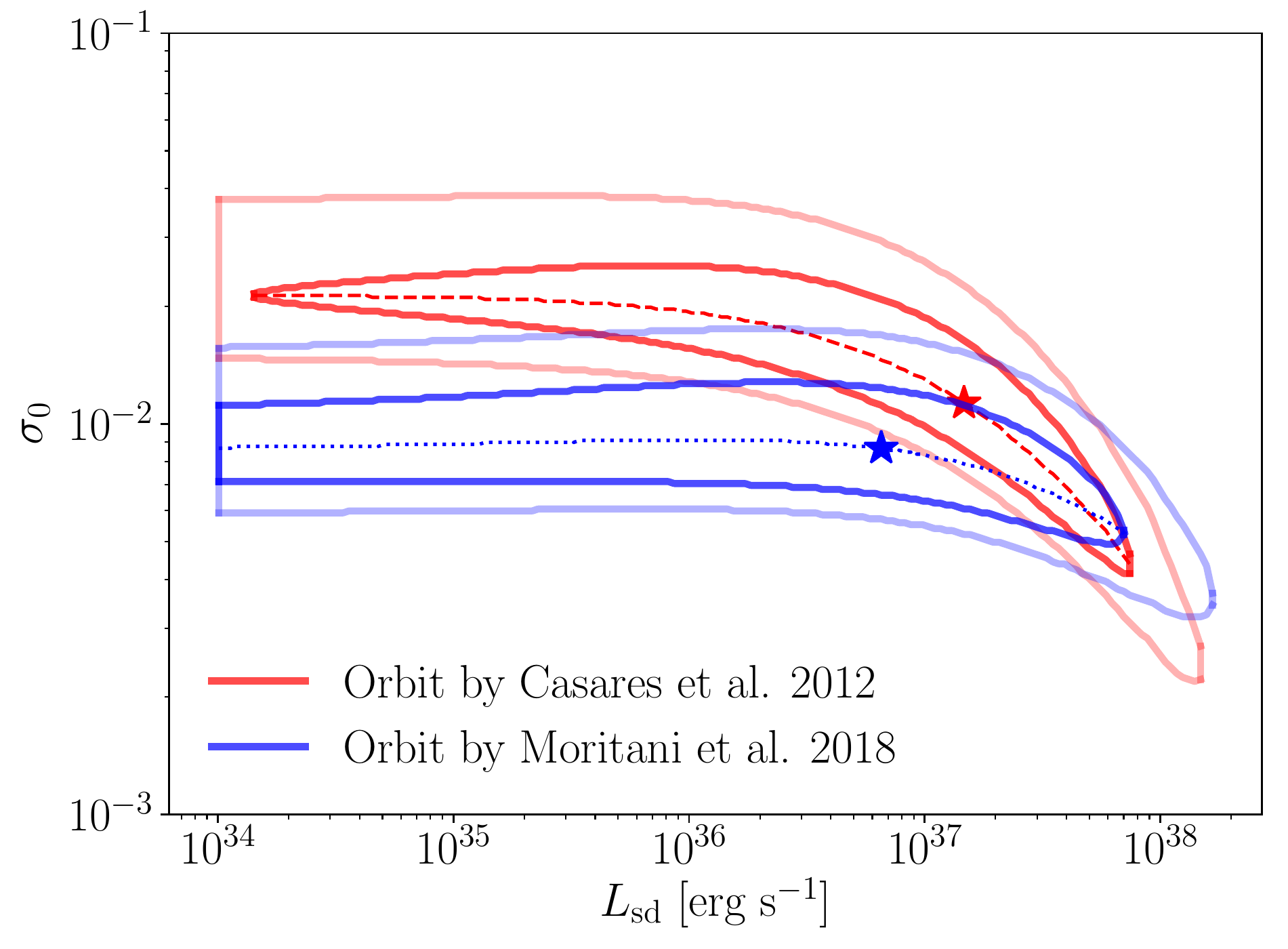}
\put(55,62.5){\icrc}
\end{overpic}
\caption{Left: Illustration of the orbit of the compact object projected onto the orbital plane for both orbital solutions. The locations of the compact object during the two sets of observations are indicated as black markers and the circumstellar disk radius is indicated by a dashed black line~\cite{Moritani-2015, Zamanov-2016}. Right: Results of the model fitting in the $\edot-\sigma_0$ plane for both sets of orbital parameters. The best solution is indicated by a star while the 1 and $2\sigma$ regions are indicated by the darker and lighter continuous lines, respectively.}
\label{fig:orbits-solutions}
\end{figure}

We summarize in this section the two available orbital solutions and the system parameters relevant for the model fitting. The properties of the companion Be star \bestar (=HD 259440) were first derived through optical spectroscopy in Ref.~\cite{Aragona-2010}. Based on that, we assume $\tbe = 30$ kK, $\rbe=7.8$~\rsun and $d=1.4$ kpc. The mass of the star \mbe is allowed to vary within the derived range according to the mass function $f(M)$ obtained in the orbital solutions and the system's inclination $i$. The orbital solution was first determined by Casares {\it et al.}~\cite{Casares-2012} through spectroscopic studies of H$\alpha$ emission lines. Recently, a totally distinct solution was proposed by Moritani {\it et al.}~\cite{Moritani-2018}, obtained with the same methodology and a larger dataset. We will use in our model fitting both solutions. The orbital period will be set to $\porb=315$ days following the results of Ref.~\cite{Aliu-2014}. The orbit of the compact object for both orbital solutions are illustrated in Fig.~\ref{fig:orbits-solutions} (left).

The compact object is assumed to be a pulsar with $\mpsr=1.4$ \msun. Under this assumption, the mass function $f(M)$ of a given orbital solution gives the relation between \mbe and the inclination $i$. Therefore, the range of $i$ can be defined by imposing that \mbe is consistent with the range derived in Ref.~\cite{Aragona-2010}. We found that $i$ is between 32 and 42\dg for the Moritani {\it et al.}~\cite{Moritani-2018} solution and that there is no value of $i$ allowed for the Casares {\it et al.}~\cite{Casares-2012} solution assuming the nominal value of $f(M)$. Thus, in order to define a set of orbital parameters based on the Casares {\it et al.}~\cite{Casares-2012} solution, we set $f(M)$ to its lower limit within the quoted uncertainties and, by doing this, we find that $i>59\dg$. In our study, $i$ will be taken as the center of the allowed range.

The properties of the stellar winds are also relevant ingredients for our model based on shocked winds. The stellar wind in Be stars is commonly described as being composed of a fast low-density polar wind and a slow dense equatorial disk wind~\cite{Waters-1988}. The properties of the disk in \bestar were studied in Refs.~\cite{Moritani-2015, Zamanov-2016} through optical spectroscopy. The disk size was estimated to be $0.85-1.4$~AU, in which the average disk radius of $1.12$~AU is indicated by dotted lines in Fig.~\ref{fig:orbits-solutions} (left), assuming the disk and the orbit of the compact object are co-planar. The orbital solutions imply that the distance between the stars is substantially larger than the disk size. Therefore, the effect of the disk on the shock formation will be neglected. The velocity of the polar wind is assumed to be $\vw = 1500$~km/s~\cite{Waters-1988}. The mass loss rate \mdot of the wind in Be stars is only poorly constrained and it is commonly assumed to be in the range $10^{-9}-10^{-8}$~\msunyr~\cite{Snow-1981,Waters-1988}. Thus, we will adopt $\mdot=10^{-8.5}$~\msunyr as a reference value and the range $10^{-9}-10^{-8}$~\msunyr will be used to provide an estimation of its uncertainties. 

\vspace{-3mm}

\section{Description of the Model}
\label{sec:model}

\vspace{-2mm}

We present here a model based on the assumption that the compact object is a pulsar. The pulsar wind is terminated due to the collision with the stellar wind and the termination shock is assumed to be the acceleration site of electron pairs from the pulsar wind. The non-thermal radiation emitted by these accelerated electrons located at the apex of the shock produces the observed X-ray and gamma-ray photons. While the X-ray photons are produced by synchrotron radiation, inverse Compton scattering (ICS) of stellar photons is responsible for producing the gamma-ray photons. The pulsar spin-down luminosity, \edot, and the magnetization of the pulsar wind, $\sigma$, are central parameters of the model.

From the hydrodynamic balance between the pulsar and the stellar wind, the distance of the shock apex to the pulsar is given by $\rsh = \frac{\sqrt{\eta}}{1+\sqrt{\eta}} D$, where $\eta = \frac{\edot}{\dot{M}v_{\mathrm{w}}c}$, and $D$ is the distance between the stars~\cite{Harding-1990, Tavani-1997, Ball-2000}. The electron pairs from the pulsar wind are accelerated at the termination shock of the pulsar wind, forming the high-energy electron population upstream of the shock which is responsible for the non-thermal radiation. The $B$-field upstream of the shock is given by $B = \sqrt{ \frac{\edot\sigma}{\rsh c (1+\sigma)} \left( 1 + \frac{1}{u^2}\right)}$, where $u$ is the radial four-velocity of the wind downstream from the shock~\cite{Kennel-1984a, Kennel-1984b}.

The electron pairs from the pulsar wind are assumed to be accelerated to a power-law energy distribution in the termination shock. After being injected into the downstream post-shock flow, radiative energy losses may change the electron energy spectrum, creating features that depart from the original power-law shape. Since the energy of the relevant electrons that describe our observations extends through a small energy range, from $\approx0.1$ to $\approx5$ TeV, its spectrum will be assumed to follow a single power-law shape of the form $\text{d}N_e/\text{d}E_e = \norm \left(E_e/1\;\text{TeV} \right)^{\slope}$. 

The seed photon field for the ICS is composed of the thermal photons radiated by the companion star. We also account for the anisotropic nature of the ICS by calculating the photon scattering angle (\thetaic) from the geometry given by the orbital solutions. Due to the relatively dense field provided by the stellar photons, the effect of pair-production absorption of gamma-rays~\cite{Gould-1967} cannot be neglected. The optical depth  of pair production absorption ($\tau_{\gamma\gamma}$) is computed and accounted for in each evaluation of the model~\cite{Sushch-2017}. The expected gamma-ray spectrum to be observed is then attenuated by a factor of $e^{-\tau_{\gamma\gamma}}$.

The present model was fitted to the SED derived from both \nustar and VERITAS observations. The two observation sets will be labelled by the lower indices $0$ and $1$, for the November and December 2017 observations, respectively. The distance between the stars ($D$) and the ICS angle (\thetaic) needed to evaluate the model were computed for a given orbital solution. The slopes of the electron spectrum (\slopei{0} and \slopei{1}) were fixed to the values derived from the single power-law fit of the X-ray spectrum (see Sec.~\ref{sec:nustar}). The free parameters of the fit are \edot, the pulsar-wind magnetization at the location of the shock $\sigma_0$, and the normalization of the electron spectrum for both periods, \normi{0} and \normi{1}. The Naima package~\cite{Zabalza-2015} was used to compute the synchrotron and ICS emission. The SED fit was performed by a $\chi^2$ method using the Minuit framework~\cite{James-1975}. 

\vspace{-3mm}

\section{Results}
\label{sec:results}

\vspace{-2mm}

\begin{figure}[t!]
\centering
\begin{overpic}[clip, width=0.9\textwidth]{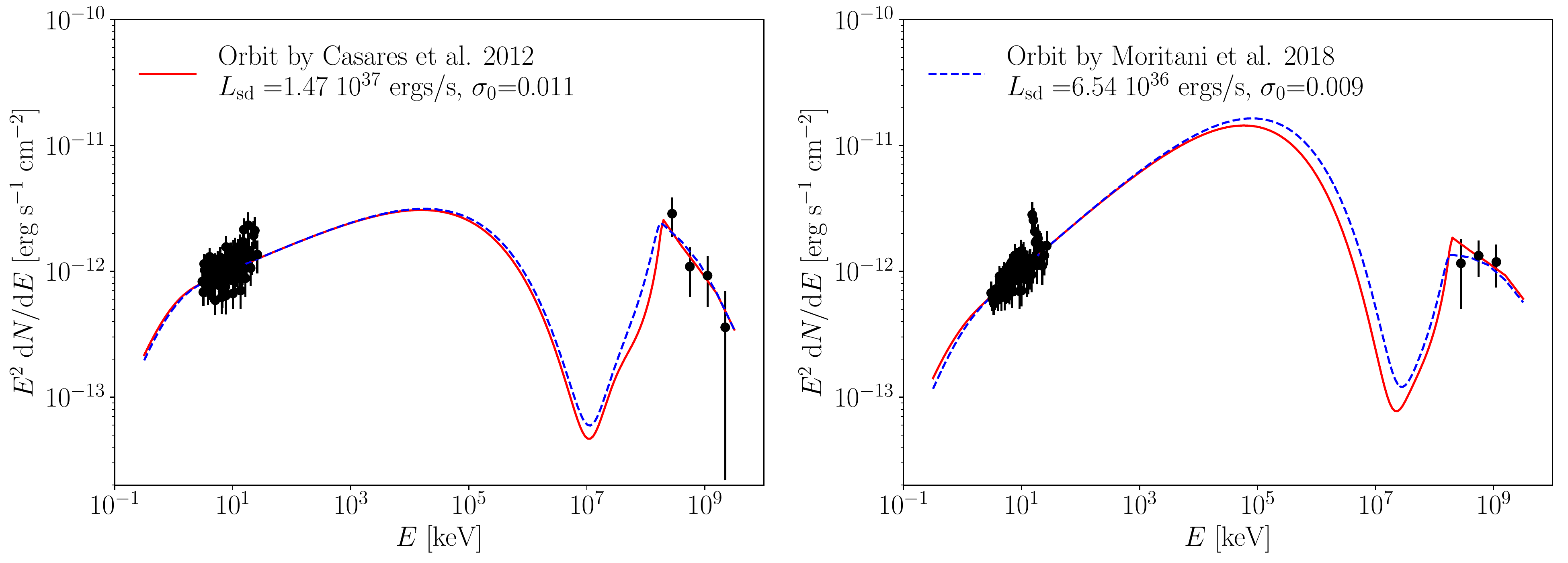}
\put(15, 7){\icrc}
\put(65, 7){\icrc}
\end{overpic}
\caption{SED data-model comparison assuming the best solution of the model fitting for both sets of orbital parameters. The November and December 2017 observations are shown in the left and right panels, respectively.}
\label{fig:sed}
\end{figure}

In Fig.~\ref{fig:orbits-solutions} (right) we show the results of the model fitting for the \edot-$\sigma_0$ plane, where the 1 and 2$\sigma$ regions for both orbital solutions are indicated. The best solutions are represented by stars and the corresponding $\chi^2/\text{dof}$ is $0.786$ for both sets of orbital parameters. The dashed lines show the $\sigma_0$ that minimizes the $\chi^2$ as a function of \edot. The SEDs are shown in Fig.~\ref{fig:sed}, where the model was evaluated for the best solution of the fit. The electron spectrum was assumed to be a power-law, starting at $E_{\text{min}}=0.2$ TeV, with an exponential cutoff characterized by $E_{\text{cut}}=5$ TeV. The values of $E_{\text{min}}$ and $E_{\text{max}}$ have no impact on the model fitting.

The observed behavior of \edot-$\sigma_0$ implies that neither of these parameters can be individually constrained with our approach. However, these results demonstrate that our observations are consistently described by a pulsar-wind model and indicate the region of the $\edot$-$\sigma_0$ space that allows for it. The impact of different system geometries is shown to be minimal, which means that both orbital parameters provide equally good solutions.

At the high \edot regime of the solutions, the $1\sigma$ upper limit is $\edot<\sn{7}{37}$ \lum for both sets of orbital parameters, which is consistent with expectations for very young pulsars. At this regime, the termination shock is closer to the companion star because of the relatively strong pulsar wind. For both orbital solutions, the termination shock is about halfway between the stars for the highest \edot values. As the \edot of the solutions decreases, the termination shock moves closer to the pulsar, within a small fraction of the distance between the stars. At this regime, the $B$-field also decreases. 

The impact of the uncertainties on the system's parameters were evaluated by varying the assumed values and repeating the fitting procedure. It was found that \mdot is by far the most relevant source of uncertainty. For all the remaining system parameters the impact of their uncertainties is smaller or of the same order of the $1\sigma$ statistical uncertainty from the model fitting.   

\vspace{-3mm}

\section{Summary and Conclusions}
\label{sec:conclusions}

\vspace{-2mm}

We presented the results of two sets of combined observations of the gamma-ray binary \src, by \nustar, in the hard X-ray band, and by VERITAS, in the TeV gamma-ray band. The observations correspond to the rise of the first outburst observed in the X-ray and TeV light curve, at phases approximately $0.22$ and $0.30$. The spectral analysis performed on the \nustar observations show that the spectra are well described by a single power law model with a significant hardening observed between the two observations ($\Gamma$ going from $1.77\pm0.05$ to $1.56\pm0.05$). The SED derived from the observations was used to probe a model based on the pulsar-wind scenario. Within this model, the non-thermal emission is produced by high-energy electrons accelerated at the termination shock created by the interaction between pulsar and stellar wind.

The results of the model fitting show the regions of the $\edot$-$\sigma$ plane that are allowed by our data (see Fig.\ref{fig:orbits-solutions}, right). The $\sigma$ parameter is constrained to be $0.003-0.03$ at the location of the shock. Constraints on $\sigma$ are particularly relevant for understanding the physical process behind the transport of energy from the rotation-powered pulsar to the surrounding medium, which is still a subject of intense discussions~\cite{Arons-2002, Kirk-2009}. Theoretical models predict that at the light cylinder the pulsar wind is dominated by Poynting energy ($\sigma_{L}\gg1$), while observations of the Crab Nebula constrain $\sigma$ at much larger distances to be kinetic particle dominated ($\sigma_{N}\ll1$). The transition between these two regimes is not well described within the current theoretical framework, originating the so-called ``$\sigma$ problem''. In gamma-ray binaries, the pulsar wind termination is typically located at intermediate distances between the light cylinder and the termination shock in a pulsar wind nebulae ($\rsh\approx10^{13}-10^{14}$ cm). 
 
\vspace{-3mm}

\section*{Acknowledgments}

\vspace{-2mm}

{
\footnotesize

This work used data from the \nustar mission, a project led by the California Institute of Technology, managed by the Jet Propulsion Laboratory, and funded by NASA. We made use of the \nustar Data Analysis Software (NuSTARDAS) jointly developed by the ASI Science Data Center (ASDC, Italy) and the California Institute of Technology (USA).

VERITAS is supported by grants from the U.S. Department of Energy Office of Science, the U.S. National Science Foundation and the Smithsonian Institution, and by NSERC in Canada. This research used resources provided by the Open Science Grid, which is supported by the National Science Foundation and the U.S. Department of Energy's Office of Science, and resources of the National Energy Research Scientific Computing Center (NERSC), a U.S. Department of Energy Office of Science User Facility operated under Contract No. DE-AC02-05CH11231. We acknowledge the excellent work of the technical support staff at the Fred Lawrence Whipple Observatory and at the collaborating institutions in the construction and operation of the instrument.

}

\end{document}